\documentclass[a4paper,11pt]{article}
\pdfoutput=1 

\usepackage{jcappub} 

\usepackage[T1]{fontenc} 
\usepackage{graphicx}
\usepackage{dcolumn}
\usepackage{bm}
\usepackage{amsmath,multirow}
\usepackage{amssymb}
\usepackage{epstopdf,tabularx}

\title{\boldmath Generalized flow-composed symplectic methods
for post-Newtonian Hamiltonian systems}

\author[a]{Shixiang Huang,}
\author[b,c]{Kaiming Zeng,}
\author[b,c]{Xinghua Niu}
\author[b,c,1]{and Lijie Mei\note{Corresponding author.}}

\affiliation[a]{School of Mathematics \& Computational Science,
Shangrao Normal University, Shangrao 334001, China}
\affiliation[b]{School of Mathematics, Yunnan Normal
University, Kunming 650500, China}
\affiliation[c]{Yunnan Key Laboratory of Modern Analytical Mathematics
and Applications, Yunnan Normal University, Kunming 650500, China}

\emailAdd{hsx8154562@126.com}
\emailAdd{15907536773@163.com}
\emailAdd{1945326539@qq.com}
\emailAdd{bxhanm@126.com}

\abstract{
Due to the nonseparability of the post-Newtonian (PN) Hamiltonian
systems of compact objects, the symplectic methods that admit the
linear error growth and the near preservation of first integrals are always
implicit as explicit symplectic methods have not been currently found for general
nonseparable Hamiltonian systems. Since the PN Hamiltonian has a particular
formulation that includes a dominant Newtonian part and a perturbation
PN part, we present the generalized flow-composed
Runge--Kutta (GFCRK) method with a free parameter $\lambda$
to PN Hamiltonian systems. It is shown that the
GFCRK method is symplectic once the underlying RK method is symplectic,
and it is symmetric once the underlying RK method is symmetric under the
setting $\lambda=1/2$. Numerical experiments with the 2PN Hamiltonian
of spinning compact binaries demonstrate the higher accuracy and efficiency
of the symplectic GFCRK method than the underlying symplectic RK
method in the case of weak PN effect. Meanwhile,
the numerical results also support  higher efficiency of the symplectic GFCRK
method than the semi-explicit mixed symplectic method of the same order.}

\begin{document}
\maketitle
\flushbottom

\section{Introduction}

Post-Newtonian (PN) Hamiltonian approach has been verified
to be a good weak-filed approximation to general relativity.
It is usually used to model the system of compact objects
in the early stage of the inspiral \cite{Baker2007}.
Except for some qualitative results
on the dynamics of PN Hamiltonian systems
\cite{Wu2011,Wu2015,Wu2015b}, the numerical simulation
involving the design and implementation of numerical integration
methods has been necessary for PN Hamiltonian systems.

Numerous numerical methods have been developed to solve
the PN Hamiltonian system numerically. The generic designing methods
mainly include the single-step method such as the classical
Runge--Kutta (RK) method and linear multistep methods, and
high-order methods are often employed for short-term orbit prediction \cite{Hairer1987}.
However, for long-term dynamical investigation, these generic
designing methods always failed to obtain reliable numerical
solutions due to the extremely enlarged truncation errors, which
are caused by the at least quadratic growth of global
errors with respect to the integration time \cite{Calvo1995}.

The development of symplectic methods \cite{Hairer2006,Feng2010}
takes a great advantage
to the numerical solving of Hamiltonian systems, since symplectic
methods admit a linear error growth for near-integrable Hamiltonian
systems and could nearly preserve the first integrals of the Hamiltonian.
Up to now, symplectic methods have been well developed for
Hamiltonian problems. Except for the generic symplectic methods
for general Hamiltonian systems, such as the symplectic Runge--Kutta \cite{ Sanz-Serna1988}
method and the generating function method \cite{Feng2010}, typical explicit
symplectic methods for Newtonian N-body problems include the standard
$T+V$ splitting methods \citep{Ruth1983,Forest1990},
the $H_0+\varepsilon H_1$ splitting methods \citep{Wisdom1991},
the force gradient methods \citep{Chin1997,Chin2007,Omelyan2002},
and the pseudo-high-order methods
\citep{McLachlan1995,Chambers2000,Laskar2001,Farres2013} by taking
account of the separability of the Hamiltonian.

However, for the post-Newtonian N-body problems, the Hamiltonian
is always nonseparable even though sometimes it may be integrable
in the case where the spin-spin effect is not included in the
Hamiltonian \cite{Wu2015}. Once the PN Hamiltonian is nonseparable, there
is no explicit symplectic method, and then implicit
symplectic methods are necessary for the accurate long-term
investigation. Although the fully implicit symplectic
Runge--Kutta or partitioned Runge--Kutta methods \cite{Hairer2006}
are available, their computational efficiency are low due to
the iteration used during the implementation. Instead,
Lubich et al. \cite{Lubich2010}, Zhong et al. \cite{Zhong2010},
and Mei et al. \cite{Mei2013a} proposed the mixed symplectic method that
exact solves the integrable Newtonian part and numerically
solves the perturbed PN part by the symplectic implicit midpoint
method. It is shown that the mixed symplectic method is more
efficient than the same-order fully implicit symplectic RK-type
method once the perturbation parameter is in a small
magnitude, i.e., a weak PN effect.

Because of the use of iterations in the implementation,
the mixed symplectic method is essentially implicit.
Recently, Mei \& Huang \cite{Mei2024} improved the mixed symplectic method
to the explicit near-symplectic method by replacing the
implicit midpoint method with classical explicit RK methods
to solve the PN perturbation part. Although the short-term behavior
of the improved explicit near-symplectic method is nearly
the same as the underlying mixed symplectic method on the presence
of the small perturbation parameter, the time interval on which
the linear error growth holds will be much shorter than the original
mixed symplectic method.

We also note that although the explicit \citep{Pihajoki2015,Tao2016,Liu2016}
or the implicit \cite{Jayawardana2023} extended phase space methods
perform well in the energy error in the numerical simulation
of PN Hamiltonian systems, the symplecticitiy of these explicit methods in
the original phase space and thus the most important property of
linear error growth is indefinite. It should be emphasized that
the linear error growth of the explicit extended phase space method made
by Tao \cite{Tao2016} just holds under the assumption that the extended Hamiltonian
system is near-integrable or integrable. However, as indicated by
the simple example of \cite{Tao2016}, a completely integrable Hamiltonian
could lead to a nonintegrable extended Hamiltonian, we thus consider
the near-integrability of extended Hamiltonian is hardly satisfied.

Besides the above-mentioned symplectic methods,
Anto\~{n}ana,  Makazaga \& Murua \cite{Antonana2017} proposed a
new class of symplectic methods for the Newtonian N-Body problem.
This new symplectic method is referred to as symplectic
flow-composed Runge--Kutta (FCRK) method, where the
phase flow of the Kepler two-body problem is employed to transform
the original system into a flow-composed Hamiltonian system. Since the
Kepler flow of the two-body problem dominates the evolution of the system,
the FCRK method shows much higher accuracy in the numerical simulation of
the N-body problem of our Solar System. Motivated by the FCRK method,
in this paper, we aim to adapt the symplectic FCRK method to the PN
Hamiltonian system. We find that the FCRK method could be generalized
with a free parameter $\lambda$, where $\lambda=1/2$ just reduces to
the original FCRK method proposed in \cite{Antonana2017}.
A preliminary analysis shows that
the setting $\lambda=0$ and $\lambda=1$ could slightly decrease the
computation amount in theory. Moreover, the complete freedom of $\lambda$
enables us to minimize the truncation error in principle or to preserve
the energy or other first integrals.

This paper is organized as follows.
In Section \ref{sec:formula}, we formulate the second-order PN Hamiltonian
of spinning compact binaries in the canonically conjugate variables.
In Section \ref{sec:FCRK}, we derive the GFCRK method and
present a detailed discussion on the property and its relation with the
mixed symplectic method.  Numerical experiments are presented
in Section~\ref{sec:numerical} by showing the global errors, the energy errors,
and the consumed CPU time to verify the convergence order and the
high efficiency of the GFCRK method. We draw our conclusions
in the last section.

\section{Post-Newtonian Hamiltonian formulation}\label{sec:formula}
Since the PN terms are in a smaller magnitude than the
Newtonian term of the N-body problem, the PN Hamiltonian of
the N-body problem is in a uniform formulation with the dominant
Newtonian part and a PN perturbation part
\citep{Damour2001,Andrade2001,Levi2014,Huang2014,
Dubeibe2017,HuangLi2018,Spyrou1975,Quinn1991,Chu2009}.
Without loss of generality,
in this paper we only take the conservative PN Hamiltonian of spinning
compact binaries as an example. The application of the GFCRK
method to other PN Hamiltonian systems could be extended similarly.

The PN Hamiltonian of spinning compact binaries considered in this paper
is accurate up to 2PN order. Once formulated in the Arnowitt-Deser-Misner (ADM)
coordinates and in the center-of-mass frame, the Hamiltonian reads
\begin{equation}\label{eq:Hamilton}
\begin{aligned}
H(\bm{Q},\bm{P},\bm{S}_1,\bm{S}_2) =
H_{N} + \frac{1}{c^2}H_{1PN} + \frac{1}{c^4}H_{2PN}
+\frac{1}{c^3}H_{1.5PN}^{SO}
+ \frac{1}{c^4}H_{2PN}^{SS},
\end{aligned}
\end{equation}
where $\bm{P}$ is the momenta of body 1 relative to the center,
$\bm{Q}$ is the position coordinates of body 1 relative to body 2,
$\bm{S}_i~(i=1,2)$ are the spins of the two compact bodies,
$H_{N}$, $H_{1PN}$, and $H_{2PN}$ are respectively the Newtonian
term, the 1PN-, and 2PN-order orbital contributions, $H_{1.5PN}^{SO}$
is the spin-orbit couplings of 1.5PN order, and $H_{SS}$ is the spin-spin
couplings and accurate up to 2PN order.

Let $\bm{N}=\bm{Q}/r$ be the unit vector, $r=|\bm{Q}|=\sqrt{q_1^2+q_2^2+q_3^2}$,
$m_1$ and $m_2~(m_1\leq m_2)$ be mass of the two compact bodies, $M=m_1+m_2$
be the total mass, $\beta=m_1/m_2$ be the mass ratio,
and $\eta=\beta/(1+\beta)^{2}$.
The orbital terms could be expressed as follows \cite{Buonanno2006}
\begin{eqnarray*}
  &H_{N} &= \frac{\bm{P}^2}{2}-\frac{1}{r} \\
  &H_{1PN}& =\frac{1}{8}(3\eta-1)\bm{P}^4-\frac{1}{2}\big[(3+\eta)\bm{P}^2
+\eta(\bm{N}\cdot\bm{P})^2\big]\frac{1}{r}
+\frac{1}{2r^{2}}, \\
  &H_{2PN} &=
\frac{1}{16}(1-5\eta+5\eta^2)\bm{P}^6
+\frac{1}{8}\big[(5-20\eta-3\eta^2)\bm{P}^4
-2\eta^2{(\bm{N}\cdot\bm{P})^2}\bm{P}^2
\\
&&\quad-3\eta^2{(\bm{N}\cdot\bm{P})^4}\big]\frac{1}{r}
+\frac{1}{2}\big[(5+8\eta)\bm{P}^2
+3\eta (\bm{N}\cdot\bm{P})^2\big]\frac{1}{r^2}
-\frac{1}{4}(1+3\eta)\frac{1}{r^3}.
\end{eqnarray*}
Moreover, the spin-orbit \cite{Nagar2011} and spin-spin
\cite{Buonanno2006} terms are respectively:
\begin{eqnarray*}
  &H_{1.5PN}^{SO} &= \frac{1}{r^{3}} \big(2\bm{S}
+\frac{3}{2}\bm{S}^{*}\big)\cdot\bm{L}, \\
  &H_{2PN}^{SS}&=\frac{1}{2r^3}
\big[3(\bm{S}_0\cdot\bm{N})^2-\bm{S}_0^2\big],
\end{eqnarray*}
where $\bm{S}= \bm{S}_1 + \bm{S}_2$,
$\bm{S}^{*}=\frac{1}{\beta}\bm{S}_1+\beta\bm{S}_2$,
$\bm{S}_{0} = \bm{S} + \bm{S}^{*}$, and
$\bm{L}$ is the orbital angular momentum vector
$\bm{L}=\bm{Q} \times \bm{P}$.

For the gravitational constant $G$ and the total
mass $M$, we employ the convenient geometric unit $G=M=1$.
As explained in \cite{Lhotka2015,Dubeibe2017,HuangLi2018,HuangLi2019},
the rescaled speed of light $c$ should be readjusted to
coincide with specific physical phenomena under the setting $G=M=1$.
A smaller value of $c$ for \eqref{eq:Hamilton} usually
indicates a stronger PN effect. The details on the discussion
of $c$ can be found in \cite{HuangLi2018,HuangLi2019}.

Because the evolution equations of \eqref{eq:Hamilton} read
\begin{equation}\label{eq:evolution}
\begin{aligned}
\frac{\mathrm{d}\bm{Q}}{\mathrm{d}t}=\frac{\partial {H}}{\partial \bm{P}},
\qquad \frac{\mathrm{d}\bm{P}}{\mathrm{d}t}=-\frac{\partial {H}}{\partial \bm{Q}},
\qquad \frac{\mathrm{d}\bm{S}_i}{\mathrm{d}t}=\frac{\partial {H}}{\partial
\bm{S}_i}\times \bm{S}_i,~~ i=1,2,
\end{aligned}
\end{equation}
the spin variables $\bm{S}_i$ are not canonically conjugate
to each other. On noting the conservation of the spin magnitudes
$|\bm{S}_i|$ during the evolution of the system, Wu \& Xie \cite{Wu2010}
introduced the canonical conjugate spin variables
$\bm{\theta}=(\theta_1,\theta_2)$ and $\bm{\xi}=(\xi_1,\xi_2)$:
\begin{equation}\label{spin-new-v}
\bm{S}_i=\left(
\begin{aligned}
&\rho_{i}\cos(\theta_{i})
\\
&\rho_{i}\sin(\theta_{i})
\\
&\quad \xi_{i}
\end{aligned}
\right),\quad i=1,2,
\end{equation}
where $\rho_{i}^2+\xi_{i}^2=\Lambda_i^2$, $\rho_{i}>0$, and $\Lambda_i=|\bm{S}_i|$
for $i=1,2$. Usually, the spin magnitudes are expressed as
$\Lambda_i=\chi_{i}m_{i}^{2}/M^{2}$ with $\chi_i\in [0, 1]$.
Because of the conservation of $\Lambda_i$, the usage of the new
canonical conjugate spin variables reduces the 12-dimensional noncanonical
Hamiltonian system $H(\bm{Q},\bm{P},\bm{S}_1,\bm{S}_2)$ to a 10-dimensional
canonical Hamiltonian system $H(\bm{Q},\bm{P},\bm{\theta},\bm{\xi})$,
where the canonical equations follow that
\begin{equation}\label{eq:evolution-canonical}
\begin{aligned}
&\frac{\mathrm{d}\bm{Q}}{\mathrm{d}t}=\frac{\partial {H}}{\partial \bm{P}},
\qquad \frac{\mathrm{d}\bm{P}}{\mathrm{d}t}=-\frac{\partial {H}}{\partial \bm{Q}},
\qquad \frac{\mathrm{d}\bm{\theta}}{\mathrm{d}t}=\frac{\partial {H}}{\partial\bm{\xi}},
\qquad \frac{\mathrm{d}\bm{\xi}}{\mathrm{d}t}=-\frac{\partial {H}}{\partial\bm{\theta}}.
\end{aligned}
\end{equation}

\section{New symplectic method with composed flows}\label{sec:FCRK}
\subsection{Generalized flow-composed method}
To describe the numerical method, we let $\bm{z} = (\bm{P},\bm{\xi},
\bm{Q},\bm{\theta})$, $I_5$ be the $5\times5$ identity matrix, and
$J=\left(
       \begin{array}{cc}
         O & I_{5} \\
         -I_{5} & O \\
       \end{array}
     \right)
$
be the canonical skew-symmetric matrix. Then, the canonical equations
corresponding to $H(\bm{z}):=H(\bm{Q},\bm{P},\bm{\theta},\bm{\xi})$
could be formally written as
\begin{equation}\label{eq:formal}
\frac{\mathrm{d}\bm{z}}{\mathrm{d}t} = J^{-1}\nabla H(\bm{z}).
\end{equation}
Since the rescaled parameter $c$ is usually in a magnitude larger
than 1, we spit the Hamiltonian \eqref{eq:Hamilton} into two
parts as follows:
\begin{equation}\label{eq:splitHamilton}
H(\bm{z}) = H_N + \varepsilon H_{PN},
\end{equation}
where $\varepsilon = \frac{1}{c^2}<1$ and
$H_{PN} = H_{1PN} + \varepsilon H_{2PN} +
\sqrt{\varepsilon}H_{1.5PN}^{SO}
+ \varepsilon H_{2PN}^{SS}$.
Using the notations $f=J^{-1}\nabla H_N$
and $g=J^{-1}\nabla H_{PN}$,
we rewrite the equation \eqref{eq:formal} as follows:
\begin{equation}\label{eq:split}
\frac{\mathrm{d}\bm{z}}{\mathrm{d}t}
= f(\bm{z}) + \varepsilon g(\bm{z}).
\end{equation}

Because $H_N$ is just the Hamiltonian of the Newtonian
two-body problem, it is completely integrable and could be
exactly solved in principle. Therefore, we assume that
there exists the phase flow $\phi_t: \bm{z}_0\mapsto
\bm{z}(t)$ such that $\bm{z}(t)$ is the solution of
the initial-value problem
\begin{equation*}
\frac{\mathrm{d}\bm{z}}{\mathrm{d}t}
= f(\bm{z}), \quad \bm{z}(0)=\bm{z}_0,
\end{equation*}
corresponding to the Hamiltonian $H_N$. Let $\tau$ be a
given constant and consider the change of variables:
\begin{equation}\label{eq:change-v}
\bm{z}(t) = \phi_{t-\tau}(\bm{w}(t)),
\end{equation}
the canonical equation \eqref{eq:split} is transformed into
\begin{equation}\label{eq:trans}
\frac{\mathrm{d}\bm{w}}{\mathrm{d}t}
=\varepsilon \Big(\frac{\partial\phi_{t-\tau}(\bm{w})}
{\partial \bm{w}}\Big)^{-1}g\big(\phi_{t-\tau}(\bm{w})\big),
\end{equation}
in the new variables $\bm{w}$.

Since $\phi_t$ is the phase flow of the Hamiltonian $H_{N}$,
the Joacobian $\frac{\partial\phi_{t-\tau}(\bm{w})}
{\partial \bm{w}}$ is symplectic and nonsingular, i.e.,
\begin{equation*}
\Big(\frac{\partial\phi_{t-\tau}(\bm{w})}
{\partial \bm{w}}\Big)
^\intercal J\Big(\frac{\partial\phi_{t-\tau}(\bm{w})}
{\partial \bm{w}}\Big)=J,
\end{equation*}
which means
\begin{equation*}
\Big(\frac{\partial\phi_{t-\tau}(\bm{w})}
{\partial \bm{w}}\Big)^{-1}J^{-1}=J^{-1}
\Big(\frac{\partial\phi_{t-\tau}(\bm{w})}
{\partial \bm{w}}\Big)
^\intercal.
\end{equation*}
On noting $g=J^{-1}\nabla H_{PN}$, we thus derive from
\eqref{eq:trans} that
\begin{equation}\label{eq:conHamil}
\frac{\mathrm{d}\bm{w}}{\mathrm{d}t}
=\varepsilon J^{-1}\Big(\frac{\partial\phi_{t-\tau}(\bm{w})}
{\partial \bm{w}}\Big)^{\intercal}\nabla H_{PN}
\big(\phi_{t-\tau}(\bm{w})\big),
\end{equation}
which is just the canonical equation of the Hamiltonian
\begin{equation}\label{eq:Hamil-newV}
\mathcal{H}(\bm{w}) = \varepsilon H_{PN}\big(\phi_{t-\tau}(\bm{w})\big).
\end{equation}

Now, we consider the constant stepsize numerical method.
Suppose that $h$ is the stepsize, the Hamiltonian is numerically
integrated on the interval $[0,T_{end}]$.
Suppose that $\tau =t_0+ \lambda h$ where $\lambda$ is
a constant. Let $t=t_0 + \zeta h$ and $\widetilde{\bm{w}}(\zeta)=\bm{w}(t_0+\zeta h)$,
we restrict the equation \eqref{eq:conHamil} on
 the typical interval $t\in[t_0,t_0+h]$
or equivalently $\zeta\in[0,1]$ as an initial-value problem:
\begin{equation}\label{eq:conHamil-typical}
\frac{\mathrm{d}\widetilde{\bm{w}}}{\mathrm{d}\zeta}
= h \varepsilon J^{-1}\Big(\frac{\partial
\phi_{\zeta h - \lambda h}(\widetilde{\bm{w}})}
{\partial \widetilde{\bm{w}}}\Big)^{\intercal}\nabla H_{PN}
\big(\phi_{\zeta h - \lambda h}(\widetilde{\bm{w}})\big),
\end{equation}
with the initial condition
\begin{equation}\label{eq:initialCond}
\widetilde{\bm{w}}(0)=\bm{w}(t_0)=\phi_{\lambda h}(\bm{z}(t_0)).
\end{equation}
Then, according to \eqref{eq:change-v} the solution
$\widetilde{\bm{w}}(\zeta)$ of \eqref{eq:conHamil-typical} satisfies
\begin{equation}\label{eq:transform}
\widetilde{\bm{w}}(\zeta) = \bm{w}(t_0+\zeta h)
= \phi_{(\lambda-\zeta) h}(\bm{z}(t_0+\zeta h)).
\end{equation}

Let $t_0=nh$, $\bm{Z}_{n}
\approx \bm{z}(nh)$,  $\bm{W}_{n}
\approx \bm{w}(nh)$ be the numerical solutions, and
\begin{equation}\label{eq:rhf}
G(\zeta,\widetilde{\bm{w}},\lambda,h)=J^{-1}\Big(\frac{\partial
\phi_{ \zeta h - \lambda h}(\widetilde{\bm{w}})}
{\partial \widetilde{\bm{w}}}\Big)^{\intercal}\nabla H_{PN}
\big(\phi_{ \zeta h - \lambda h}(\widetilde{\bm{w}})\big),
\end{equation}
for $n=0,1,\cdots,N-1$ with $N=T_{end}/h$.
According to \eqref{eq:initialCond}
and \eqref{eq:transform}, once apply an $s$-stage
Runge--Kutta (RK) method with the coefficients
$(a_{ij},b_{i},c_{i})_{i,j=1}^{s}$ to the system \eqref{eq:conHamil-typical}
with the initial condition $\widetilde{\bm{w}}(0) = \bm{W}_{n}=
\phi_{\lambda h}(\bm{Z}_n)\approx
\bm{w}(nh)$, we yield the numerical scheme as follows:
\begin{equation}\label{FCRK}
\left\{
\begin{aligned}
& \bm{W}_{n} = \phi_{\lambda h}(\bm{Z}_n),
\\
& \bm{G}_{i}=\varepsilon G\big(c_i,\bm{W}_{n}
+ h\sum_{j=1}^{s}a_{ij}\bm{G}_{j},\lambda,h\big),
~~i=1,\cdots,s,
\\
& \bm{W}_{n+1}=\bm{W}_{n} + h\sum_{i=1}^{s}b_{i}\bm{G}_{i},
\\
& \bm{Z}_{n+1} =  \phi_{(1-\lambda) h}(\bm{W}_{n+1}),
\end{aligned}
\right.
\end{equation}
which is called generalized flow-composed RK (GFCRK) method in this paper.
It is noted that the parameter $\lambda$ is free and usually
takes values on the interval $[0,1]$. Once $\lambda=1/2$, the GFCRK
method \eqref{FCRK} reduces to the standard flow-composed RK
proposed in \cite{Antonana2017}.

It is observed from the formula of $G(\zeta,\widetilde{\bm{w}},\lambda,h)$
that the evaluation of the function $G$ involves the Jacobian
of the phase flow $\phi_t(\widetilde{\bm{w}})$. Since we cannot get the
explicit expression of $\phi_t(\widetilde{\bm{w}})$, the Jacobian is also
not explicitly obtained. As recommended by \cite{Antonana2017}, one can
use the automatic differentiation technique to compute
all the partial derivatives. Moreover, dimension $36$
of the output is larger than dimension $6$ of the input for the
automatic differentiation, we consider the forward mode to be more
favorable than the reverse mode for the automatic differentiation. In fact, the phase flow
$\phi_t(\widetilde{\bm{w}})$ of the Newtonian two-body problem contains
only an iteration procedure to solve the Kepler equation,
its Jacobian could be directly expressed in an easy manner instead
of  using automatic differentiation.

\subsection{Property of GFRCK methods}\label{property}
The properties of the GFRCK method \eqref{FCRK} certainly depend on the
free parameter $\lambda$ and the underlying RK method. We first consider
the time-symmetry of the GFCRK method. Suppose that the underlying RK
method is symmetric, i.e., the coefficients $(a_{ij},b_{i},c_{i})_{i,j=1}^{s}$
satisfy the conditions \cite{Sanz-Serna1988,Hairer2006}
\begin{equation*}
\begin{aligned}
& c_{s+1-j}=1 - c_j, \quad
b_{s+1-j} = b_{j}, \quad j=1,\cdots,s,
\\
& a_{ij} + a_{s+1-i,s+1-j} = b_j, \quad i,j=1,\cdots,s,
\end{aligned}
\end{equation*}
exchanging $n+1\leftrightarrow n$ and $h\leftrightarrow -h$ yields that
the GFCRK method \eqref{FCRK} is symmetric only if
$\lambda = 1-\lambda$, i.e., $\lambda =1/2$. Once
$\lambda\neq1/2$, the GFCRK method \eqref{FCRK} is not symmetric even if
the RK method is symmetric under our consideration that
$\phi_t$ is the phase flow of the Newtonian two-body problem.

Now, we consider the symplecticity of the GFCRK method \eqref{FCRK}.
If denote the RK method with the stepsize $h$ by $\Phi_h$, then
the numerical scheme \eqref{FCRK} is abstractly expressed by
\begin{equation*}
\bm{Z}_{n+1} = \Psi_h\big(\bm{Z}_{n}\big).
\end{equation*}
where $\Psi_h=\phi_{(1-\lambda) h}\circ\Phi_h\circ\phi_{\lambda h}$.
Since $\phi_{(1-\lambda) h}$ and $\phi_{\lambda h}$ are phase flows
of the Hamiltonian $H_N$ with different stepsizes, they are both
symplectic mappings. This means that the symplecticity of
the numerical discrete flow $\Psi_h$ remains the same as that
of $\Phi_h$ on noting the fact that the composition of symplectic
mappings is still symplectic. That is, once the RK method is symplectic,
i.e., the coefficients $(a_{ij},b_{i},c_{i})_{i,j=1}^{s}$
satisfy the conditions  \cite{Hairer2006}
\begin{equation*}
b_ia_{ij} + b_ja_{ji} = b_ib_j,\quad i,j=1,\cdots,s,
\end{equation*}
the numerical discrete flow $\Psi_h$ and thus the GFCRK method \eqref{FCRK}
is symplectic.

Moreover, it is also known that the GFCRK method \eqref{FCRK}
is nonsymplectic provided the RK method is nonsymplectic.
A typical example of symplectic RK methods is the second-order
implicit midpoint method. Higher-order symplectic RK methods
mainly include the Gauss--Legendre collocation method.
Here, we present the Butcher tabular of the fourth-order Gauss--Legendre
collocation method IRK4 \cite{Hairer1987,Hairer2006} as an example:
\begin{equation*}\label{butcher-IRK4}
\renewcommand{\arraystretch}{1.5}
\begin{array}{ccc}
\begin{tabular}{ c|cc}
$c_1$    &$a_{11}$  & $a_{12}$  \\
$c_2$    & $a_{21}$ & $a_{22}$  \\
 \hline
&$\raisebox{-0.5ex}[0.5pt]{$b_1$}$
&$\raisebox{-0.5ex}[0.5pt]{$b_2$}$
\end{tabular}
&=&
\begin{tabular}{ c|cc}
$\frac{1}{2}-\frac{\sqrt{3}}{6}$   &   $\frac{1}{4}$  &
$\frac{1}{4}-\frac{\sqrt{3}}{6}$
\\
$\frac{1}{2}+\frac{\sqrt{3}}{6}$   &   $\frac{1}{4}+\frac{\sqrt{3}}{6}$
& $\frac{1}{4}$
\\
 \hline
& \raisebox{-0.5ex}[0.5pt]{$\frac{1}{2}$}
 & \raisebox{-0.5ex}[0.5pt]{$\frac{1}{2}$}
\end{tabular}
\end{array}\, .
\end{equation*}
More details on symplectic RK methods can be found in \cite{Hairer2006}.

We next consider the convergence order of the GFCRK method
\eqref{FCRK}. Suppose that the underlying RK method is of
order $\mu\in\mathbb{N}^{+}$. Under the assumption $\widetilde{\bm{w}}(0)
=\bm{W}_{n}=\bm{w}(nh)=\phi_{\lambda h}(\bm{z}(nh))$,
we use $\widetilde{\bm{W}}_{n+1}$ and $\widetilde{\bm{Z}}_{n+1}$
to denote the numerical solutions obtained by applying the RK method to
\eqref{eq:conHamil-typical}. Due to the appearance of the small parameter
$\varepsilon$ in the equation \eqref{eq:conHamil-typical}, it will
appear in the local truncation error of $\widetilde{\bm{W}}_{n+1}$
as well. We further note that since the function
$G(\zeta,\widetilde{\bm{w}},\lambda,h)$ defined in
\eqref{eq:rhf} essentially involves the Kepler phase flow $\phi_t$,
the power of $\varepsilon$ contained in the derivatives of any
order of $G(\zeta,\widetilde{\bm{w}},\lambda,h)$ with respect to
$\zeta$ is just one. This gives the local truncation error of
$\mathcal{O}(\varepsilon h^{\mu+1})$ instead of
$\mathcal{O}(\varepsilon^{\mu+1} h^{\mu+1})$ for the
numerical solutions $\widetilde{\bm{W}}_{n+1}$, i.e.,
\begin{equation}\label{localerror}
\widetilde{\bm{W}}_{n+1}-\bm{w}((n+1)h)=
\mathcal{O}(\varepsilon h^{\mu+1}).
\end{equation}

Because the Hamiltonian $H_N$
is completely integrable, the continuous dependency of
the phase flow $\phi_t$ along with \eqref{localerror}
leads to that the local error
of $\widetilde{\bm{Z}}_{n+1}$ also satisfies
\begin{equation*}\label{localerror2}
\widetilde{\bm{Z}}_{n+1}-\bm{z}((n+1)h)=
\mathcal{O}(\varepsilon h^{\mu+1}),
\end{equation*}
which means the global error of the GFCRK method satisfies
\begin{equation}\label{GlobalError}
\bm{Z}_{n}-\bm{z}(nh)=\mathcal{O}(\varepsilon h^{\mu}),
\end{equation}
for all $n=1,2,\cdots,N$.
However, the direct application of the RK method to the
canonical equation \eqref{eq:formal} could only yield
the following local error estimation
\begin{equation*}\label{localerror3}
\widetilde{\bm{Z}}_{n+1}-\bm{z}((n+1)h)=\mathcal{O}(h^{\mu+1}),
\end{equation*}
which gives the following global error estimation
\begin{equation}\label{GlobalError2}
\bm{Z}_{n}-\bm{z}(nh)=\mathcal{O}(h^{\mu}),
\end{equation}
for all $n=1,2,\cdots,N$.
That is, the global error of the GFCRK method will be
smaller than that of the underlying RK method with a factor
of $\varepsilon<1$ when applied to numerically integrate
the post-Newtonian Hamiltonian \eqref{eq:Hamilton}.

Once the GFCRK method and its underlying RK method are both
symplectic, they are implicit as well. As recommended in \cite{Hairer2006},
we use the fixed-point iteration to find the numerical solutions
of $\bm{G}_i$ for $i=1,\cdots,s$.
In this case, the iteration function of the GFCRK method becomes
\begin{equation}\label{eq:iteration1}
\mathcal{F}(\bm{G}_1,\cdots,\bm{G}_s) =
\varepsilon G(\bm{G}_1,\cdots,\bm{G}_s).
\end{equation}
However, the iteration function of the RK method directly applied
to \eqref{eq:split} will be
\begin{equation}\label{eq:iteration2}
\mathcal{U}(\bm{G}_1,\cdots,\bm{G}_s) =f(\bm{G}_1,\cdots,\bm{G}_s)
+\varepsilon g(\bm{G}_1,\cdots,\bm{G}_s).
\end{equation}
The comparison between \eqref{eq:iteration1} and
\eqref{eq:iteration2} indicates a faster convergence
of the GFCRK method than the RK method provided
$\varepsilon<1$. This point will also be illustrated
by the numerical result in the next section.

For the special case where $H_N$ is replaced by a quadratic
Hamiltonian and thus $f(\bm{z})$ is linear as a function of $\bm{z}$,
the GFCRK method \eqref{FCRK} will reduce to the Lawson's generalized
RK method, i.e., a class of exponential RK (ERK) methods. In particular,
the reduced ERK methods will be the same regardless
of the value of $\lambda$. In this case, the time-symmetry and
the symplecticity of the ERK method will be the same
as the underlying RK method. More details such as the issue of
the convergence of the fixed-point iteration for symplectic ERK methods
can be found in \cite{Lawson1967,Celledoni2008,Mei2017}.

We finally consider the dependence of the performance of GFCRK
methods on the free parameter $\lambda$.
For the case of a constant $\tau$, such as $\tau=0$ or $\tau=\lambda h$
with a fixed $\lambda$, the GFCRK method is referred to as
\emph{nonlinear Lawson's RK method} \cite{Antonana2017}, which uses the same
change of variables $\bm{z}(t) = \phi_{t-\tau}(\bm{w}(t))$ for different
time steps. However, the authors of \cite{Antonana2017} claimed that once applied to the
perturbed Kepler problem the fixed-point iteration of the \emph{nonlinear
Lawson's RK method} with a constant $\tau$ converges slower for larger values
of $T_{end}$ because the divergence between two nearby trajectories in
the perturbed Kepler problem linearly
increases with time (More precisely, the linear
increase of the divergence depends on the integrability
of the perturbed Kepler problem. If the perturbed system is still completely
integrable, the linear increase holds for all $T_{end}\in\mathbb{R}^{+}$.
However, if the perturbed problem is nearly integrable, the linear increase
holds roughly for the time interval whose length is inversely proportional to
the perturbation parameter).

In this paper, we set $\tau = t_0 + \lambda h$
with a fixed $\lambda$ and a varying $t_0$ for the GFCRK method.
Since the change of variables is $\bm{z}(t) = \phi_{t-\tau}(\bm{w}(t))$,
we actually use different change of variables for different time steps
as $\tau$ varies with $n$ by setting $t_0=nh$ for
$n=1,\cdots,N-1$ provided that the free parameter $\lambda$ is fixed.
As claimed in \cite{Antonana2017}, the case of $\lambda=1/2$ for the GFCRK method
performs much better than the \emph{nonlinear Lawson's RK method}.
Meanwhile, as previously analyzed the value of $\lambda$ introduces significant
influence not to the symplecticity, the order, and the convergence of the fixed-point
iteration, but to the symmetry of the GFCRK method as only the case of
$\lambda=1/2$ makes the method to be symmetric. This point will be illustrated
by the numerical results in the next section. Another point should be emphasized
that the setting $\lambda=0$ or $\lambda=1$ takes advantage of the unnecessary computation
of $\bm{W}_n$ or $\bm{W}_{n+1}$ in \eqref{FCRK} as
$\bm{W}_n=\phi_{0\cdot h}(\bm{Z}_n)=\bm{Z}_n$ or
$\bm{W}_{n+1}=\phi_{0\cdot h}(\bm{Z}_{n+1})=\bm{Z}_{n+1}$,
which may theoretically yield a slightly better
computational efficiency than any other values of $\lambda$.

\subsection{Relation with the mixed symplectic method}
To take advantage of the small parameter $\varepsilon$,
Lubich et al. \cite{Lubich2010}, Zhong et al. \cite{Zhong2010},
and Mei et al. \cite{Mei2013a} considered the mixed symplectic
method that mixes an explicit integrator for the completely integrable $H_N$
part with an implicit integrator for the nonintegrable nonseparable
$H_{PN}$ perturbation via the splitting approach. To describe
the mixed symplectic method, we first introduce the Lie derivative
operators
\begin{equation*}
\begin{aligned}
X=\{\,\cdot\,,H\}=J^{-1}\frac{\partial H}{\partial\bm{z}}
\frac{\partial}{\partial\bm{z}}
=\frac{\partial H}{\partial\bm{P}}\frac{\partial }{\partial\bm{Q}}
-\frac{\partial H}{\partial\bm{Q}}\frac{\partial }{\partial\bm{P}}
+\frac{\partial H}{\partial\bm{\xi}}\frac{\partial }{\partial\bm{\theta}}
-\frac{\partial H}{\partial\bm{\theta}}\frac{\partial }{\partial\bm{\xi}},
\end{aligned}
\end{equation*}
which corresponds to the canonical Hamiltonian $H$. By adopting
the splitting \eqref{eq:splitHamilton}, we denote the
Lie derivative operators corresponding to $H_N$ and $H_{PN}$
respectively by
\begin{equation*}
A=\{\,\cdot\,,H_N\}=J^{-1}\frac{\partial H_N}{\partial\bm{z}}
\frac{\partial}{\partial\bm{z}},
\end{equation*}
and
\begin{equation*}
\begin{aligned}
B=\{\,\cdot\,,H_{PN}\}=J^{-1}\frac{\partial
H_{PN}}{\partial\bm{z}}\frac{\partial}{\partial\bm{z}},
\end{aligned}
\end{equation*}
which absolutely satisfy $X=A+\varepsilon B$.

It is noted that the $H_N$ part could be exactly solved.
Here, we use the exponential map $\exp(tA)$ instead of the abstract
notation $\phi_t$ to denote the phase flow of $H_N$.
Because of the nonintegrability of $H_{PN}$, it is theoretically
unsolvable and thus we use an inexact numerical integrator,
i.e., the second-order midpoint method IRK2 to approximate
the phase flow $\exp(t\varepsilon B)$ of
$\varepsilon H_{PN}$ as follows:
\begin{equation*}
\text{IRK2}(h) = \exp\big(\varepsilon hB + \mathcal{O}(\varepsilon^3 h^3)\big).
\end{equation*}
Then, following the approach of composition method
or splitting method, we derive the mixed symplectic method:
\begin{equation*}
\text{Semi2:~~~~} \exp(\tfrac{h}{2}A)\circ\text{IRK2}(h)\circ\exp(\tfrac{h}{2}A)
=\exp\big(hX + \mathcal{O}(\varepsilon h^3)\big),
\end{equation*}
which shows the second-order accuracy of Semi2 and the appearance
of the small parameter $\varepsilon$ in the dominant truncation error.
Due to the symplecticity and symmetry of IRK2, the derived
mixed method Semi2 is certainly symplectic and symmetric for the
Hamiltonian $H$. Moreover, because IRK2 is an implicit scheme,
Semi2 also needs the fixed-point iteration to obtain numerical solutions
and therefore it is usually referred to as
\emph{semi-explicit} or \emph{semi-implicit}.

Following Yoshida's triple symmetric composition approach \cite{Yoshida1990},
we could derive the fourth-order mixed symplectic method Semi4:
\begin{equation*}
\text{Semi4~:=~Semi2}(\gamma_0 h)\circ\text{Semi2}((1-2\gamma_0) h)\circ
\text{Semi2}(\gamma_0 h),
\end{equation*}
and the sixth-order mixed symplectic method Semi6:
\begin{equation*}
\text{Semi6~:=~Semi4}(\gamma_1 h)\circ\text{Semi4}((1-2\gamma_1) h)\circ
\text{Semi4}(\gamma_1 h),
\end{equation*}
where $\gamma_0 = \frac{1}{2-2^{1/3}}$ and $\gamma_1 = \frac{1}{2-2^{1/5}}$.
Higher-order mixed symplectic methods could be successively derived.

From the construction, it is known
that both Semi4 and Semi6 take the advantage of the appearance
of the small parameter $\varepsilon$ in the truncation error,
which is the same as the symplectic GFCRK method derived in
the previous subsection. This point indicates the superior
error performance of the mixed symplectic method and the
symplectic GFCRK method over the classical fully implicit
Gauss-type symplectic method. Furthermore, if we use the
second-order implicit midpoint method, i.e.,
$a_{11}=c_1=\frac{1}{2},~c_1=1$, and set $\lambda=1/2$,
the GFRCK method \eqref{FCRK} is surprisingly identical
to the second-order mixed symplectic method Semi2.

The final point concerns computational efficiency.
It has been illustrated by numerous numerical experiments
that the mixed symplectic method is more efficient
than the same order fully implicit symplectic method
provided $\varepsilon <1$, that is the former consumes
less CPU time than the latter if a certain accuracy of
the numerical solution is prescribed. A higher efficiency
of the symplectic GFCRK method than the underlying symplectic
RK method is naturally expected. As we will see in the
next section, the symplectic GFCRK method is also more
efficient than the same-order mixed symplectic method.

\section{Numerical experiments}\label{sec:numerical}

\begin{figure*}[htb]
\centering{
\includegraphics[scale=0.65]{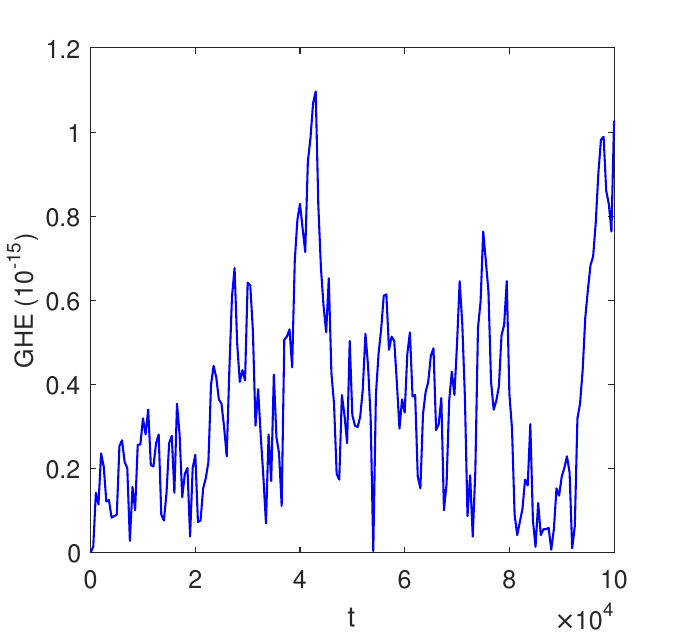}
\includegraphics[scale=0.65]{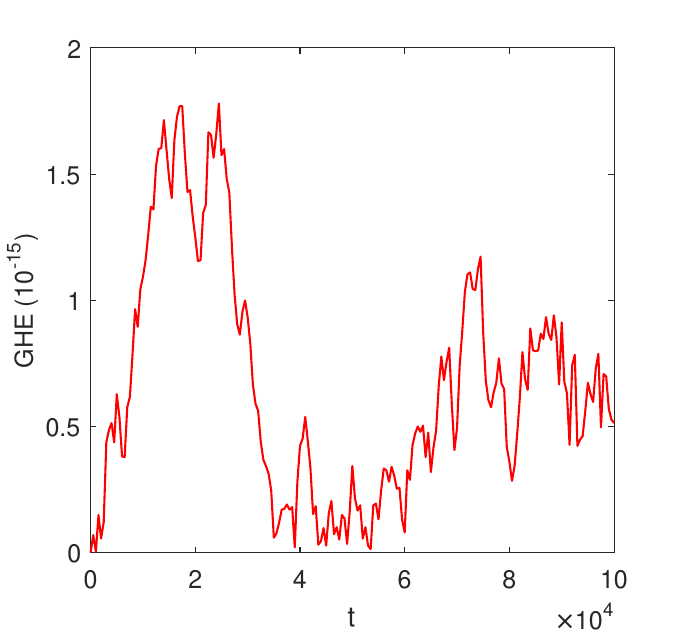}
}
\caption{\label{fig1}The global energy errors (GHE) of
the reference solutions with $c=10^{1/2}$ (left) and $c=10$ (right).}
\end{figure*}

\begin{figure*}[htb]
\centering{
\includegraphics[scale=0.65]{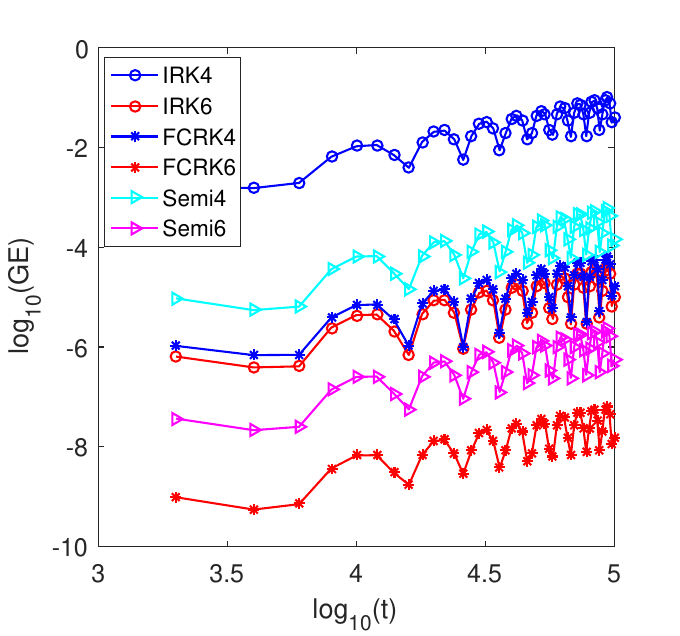}
\includegraphics[scale=0.65]{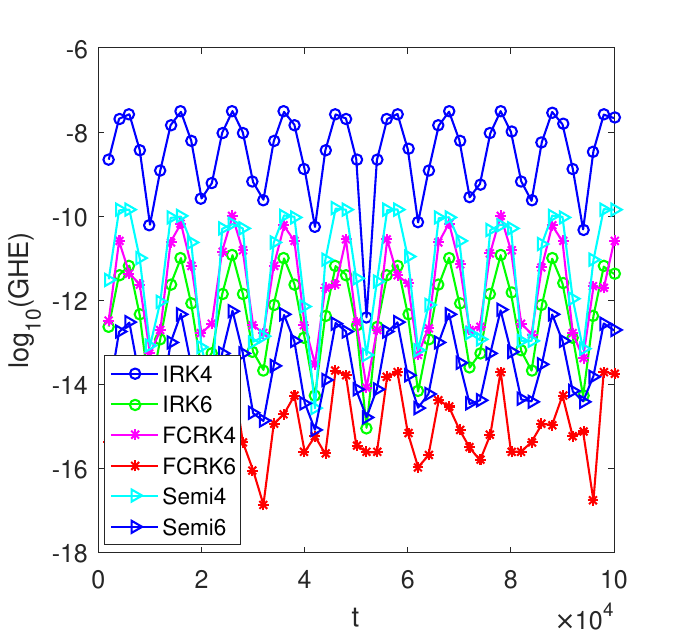}
}
\caption{\label{fig2}The global errors (GE)
and energy errors (GHE) with $h=1$ and $c=10^{1/2}$.}
\end{figure*}

\begin{figure*}[htb]
\centering{
\includegraphics[scale=0.65]{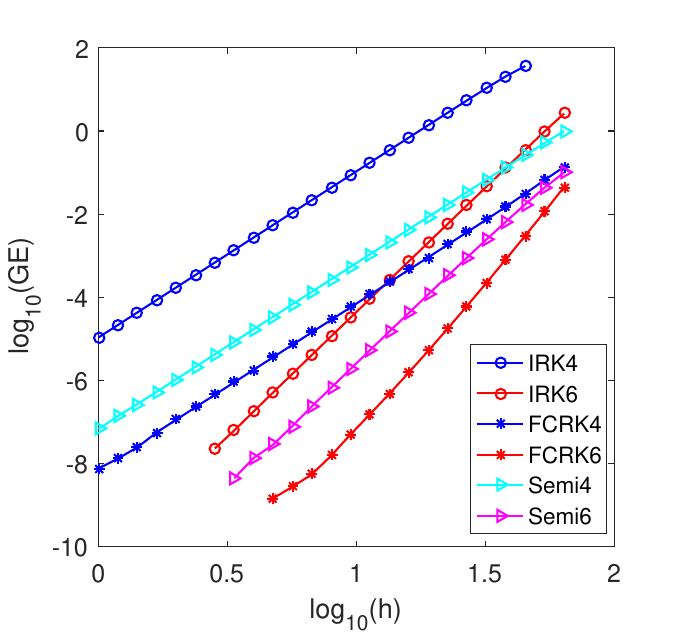}
\includegraphics[scale=0.65]{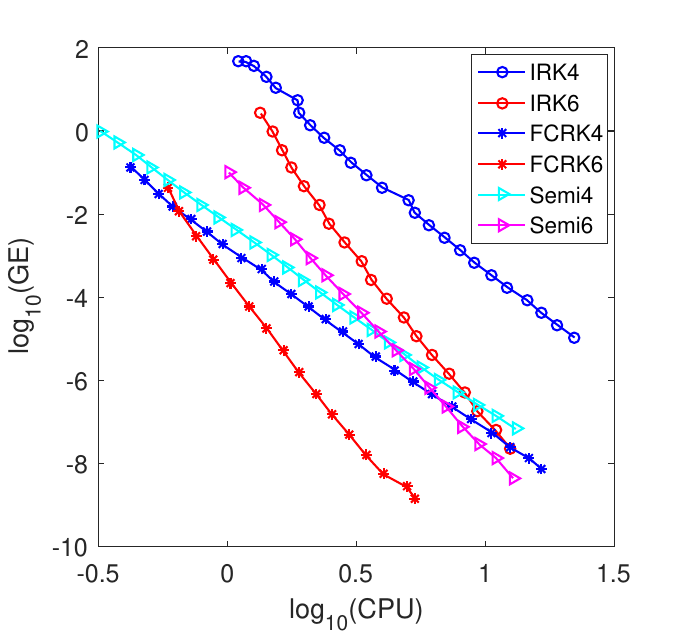}
}
\caption{\label{fig3}The numerical convergence
orders and efficiency curves with $c=10^{1/2}$.}
\end{figure*}

\begin{figure*}[htb]
\centering{
\includegraphics[scale=0.65]{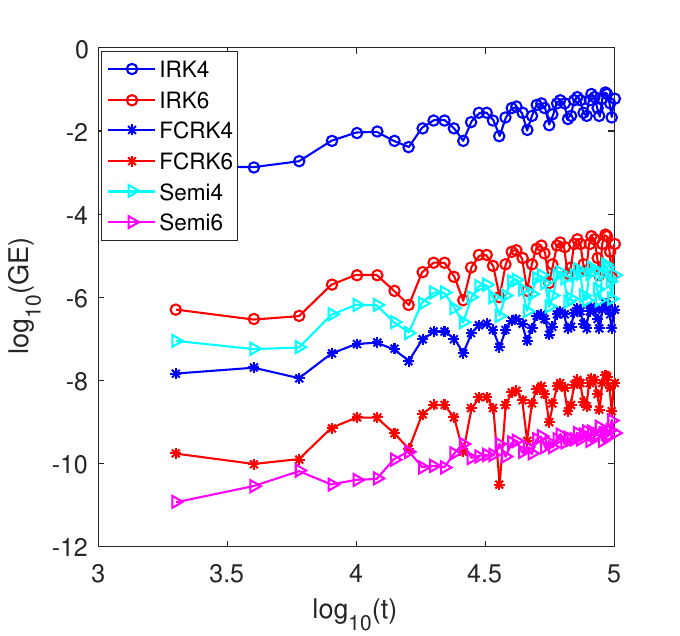}
\includegraphics[scale=0.65]{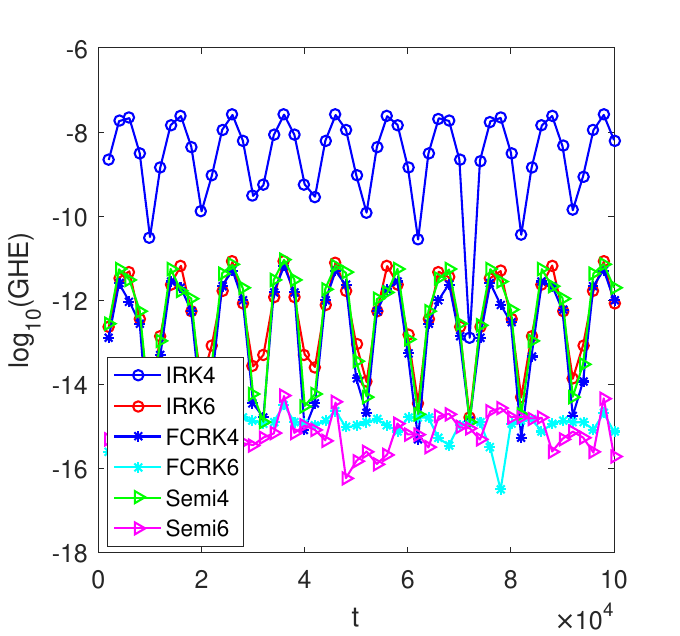}
}
\caption{\label{fig4}The global errors (GE)
and energy errors (GHE) with $h=1$ and $c=10$.}
\end{figure*}

\begin{figure*}[htb]
\centering{
\includegraphics[scale=0.65]{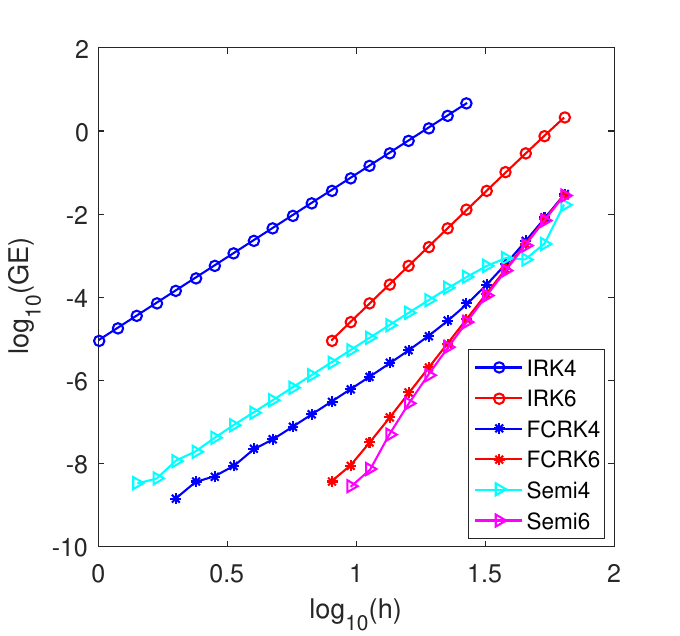}
\includegraphics[scale=0.65]{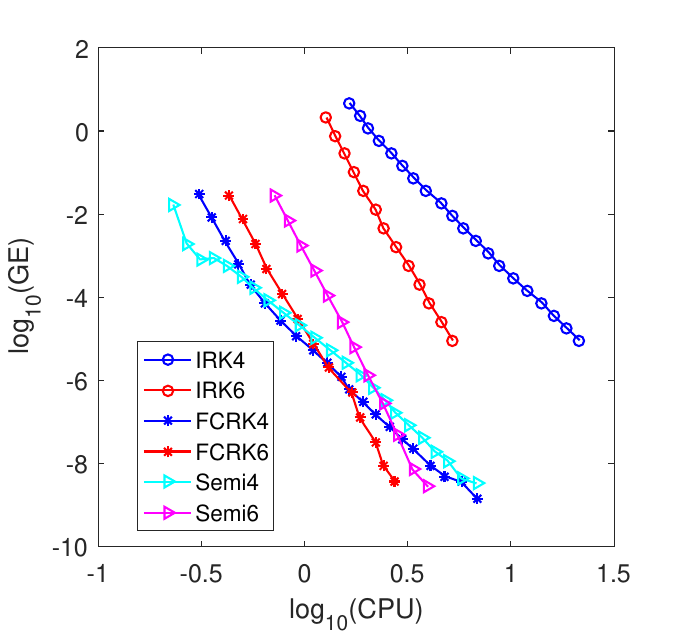}
}
\caption{\label{fig5}The numerical convergence
orders and efficiency curves with $c=10$.}
\end{figure*}

\begin{figure*}[htb]
\centering{
\includegraphics[scale=0.65]{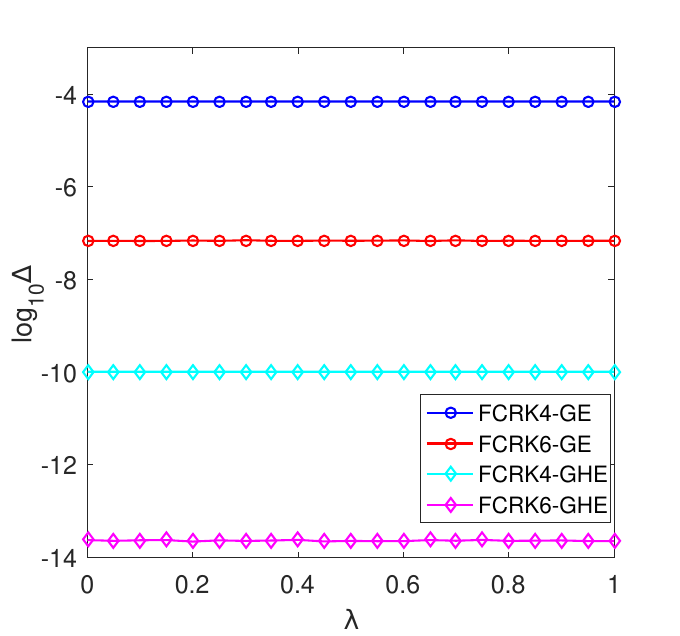}
\includegraphics[scale=0.65]{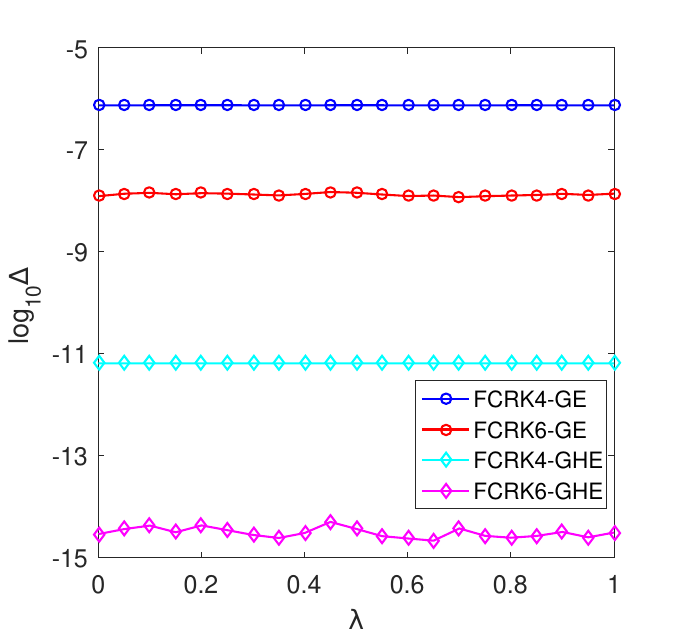}
}
\caption{\label{fig6}The dependence of global (energy)
errors on the free parameter $\lambda$ with $h=10$ and $c=10^{1/2}$ (left)
or $c=10$ (right).}
\end{figure*}

In the numerical experiments, we take the initial values as follows:
$\bm{Q}(0)=(25.34,0,0)$, $\bm{P}(0)=(0,0.18,0)$,
$\bm{\theta}(0)=(1.2490,0.6202)$,
and $\bm{\xi}(0)=(0.0445,0.0705)$.
The spin magnitudes are set as $\Lambda_1=0.0479$
and $\Lambda_2=0.6104$. The mass ratio is $\beta=0.28$.
Moreover, we use the geometric unit $G=M=1$, under which
the parameter $c$ should be readjusted to denote different
PN effects. Usually, a large value of $c$ thus a
small $\varepsilon$ indicates a weak PN effect.

In this paper, we
select six symplectic methods to make the comparison,
i.e., the classical Gauss collocation methods IRK4 and
IRK6, the symplectic GFCRK methods FCRK4 and FCRK6 respectively
based on IRK4 and IRK6, and the semi-explicit mixed symplectic
methods Semi4 and Semi6. Moreover, the free parameter $\lambda$
of the GFCRK method is fixed as $1/2$ unless otherwise stated.

In addition, we need reference solutions of very
high accuracy to measure the global error of the tested
numerical methods. Since the exact solutions cannot be
obtained in general, we regard the numerical solutions
obtained by the eighth-order Gauss symplectic collocation
method with tiny stepsize as the reference solutions.
The energy errors for the two cases $c=10^{1/2}$ and
$c=10$ are presented in Fig.~\ref{fig1}, which shows the
high accuracy of the reference solutions.

For the case of  $c=10^{1/2}$, the global errors and energy
errors of the six tested
symplectic methods are shown in Fig.~\ref{fig2}, from
which three notable points are observed. Firstly, the global
errors increase linearly with respect to time, while the
energy errors are uniformly bounded. This point clearly
coincides with the property of symplectic methods.
Secondly, the GFCRK methods FCRK4 and FCRK6 and the semi-explicit
methods Semi4 and Semi6 perform better than the fully implicit
symplectic methods, i.e., IRK4 and IRK6, which shows the
superiority of taking the advantage of the small parameter
$\varepsilon$ in the designing of numerical integrators.
Thirdly, the GFCRK methods FCRK4 and FCRK6 present a higher
accuracy than the corresponding semi-explicit methods Semi4
and Semi6.

The numerical convergence orders and efficiency curves of
the tested symplectic methods with $T_{end}= 10^5$  are shown
in Fig.~\ref{fig3}. The slopes of the curves in the left
panel of Fig.~\ref{fig3} are 3.98, 5.97, 4.02, 6.53, 3.98,
and 5.84  respectively  for IRK4, IRK6, FCRK4, FCRK6, Semi4,
and Semi6, which are in well accordance with the theoretical
convergence orders 4 and 6. The efficiency curves of global errors
versus CPU time (in seconds) shown in the right panel of Fig.~\ref{fig3}
means that the GFCRK method has the highest efficiency among
symplectic methods of the same order. We also note that the sixth-order GFCRK
method FCRK6 is most efficient among all the tested methods.

The similar results corresponding to Fig.~\ref{fig2} and Fig.~\ref{fig3}
but with $c=10$ are respectively presented in Fig.~\ref{fig4}
and Fig.~\ref{fig5}. A slight difference between Fig.~\ref{fig4}
and Fig.~\ref{fig2} is that Semi6 behaves better with a smaller
global errors than FCRK6 once $c=10$. The slops in the left
panel of Fig.~\ref{fig5} are 3.99, 5.96, 4.63, 7.29, 3.87, and
7.81 respectively  for IRK4, IRK6, FCRK4, FCRK6, Semi4, and Semi6, where
FCRK6 and Semi6 give a surprising superconvergence as the slops
are evidently larger than the theoretical value 6. Besides the
similar result that the GFCRK method of the same order is more
efficient than the semi-explicit method and the semi-explicit method
is more efficient than the RK method, it is also observed from
the right panel of Fig.~\ref{fig5} that
the high-order GFCRK/semi-explicit method intersects the
low-order GFCRK/semi-explicit method at some points, which indicates that
the high-order method is preferred under a high accuracy requirement
while the low-order method will be more efficient under a low accuracy
requirement.

A further efficiency comparison with constant stepsizes
among the six symplectic methods are
presented Table~\ref{Tab1} and Table~\ref{Tab2}, where two different
cases of $c=10^{1/2}$ and $c=10$ are tested. It is observed from the
two tables that under a constant stepsize, the Semi4 consumes the least CPU
time among all the three fourth-order method, while the CPU time
of FCRK4 is between IRK4 and Semi4. As to the sixth-order method,
FCRK6 consumes evidently less CPU time than IRK6 and Semi6 in both
the two cases of $c=10^{1/2}$ and $c=10$, while Semi6 consumes
more time than IRK6 for $c=10^{1/2}$ but less time than IRK6 for
$c=10$. That is, a small parameter $\varepsilon$ will decrease
the computational complexity of the GFCRK method and the semi-explicit
symplectic method. Overall, these two tables along with the right
panels of Fig.~\ref{fig3} and Fig.~\ref{fig5} show that although
the fourth-order FCRK4 may consume more time than the
semi-explicit Semi4 in a single stepsize, the GFCRK methods
are more efficient than semi-explicit mixed symplectic methods,
i.e., consume less CPU time under a certain accuracy requirement
for the numerical solutions.

We further investigate the dependence of the global errors or energy
errors on the free parameter $\lambda$ in Fig.~\ref{fig6}.
It is observed from this figure that the value of $\lambda$
hardly takes influence on the numerical accuracy of the
GFCRK method, since the global errors and the energy errors
nearly remain the same for different values of $\lambda$.
Moreover, we conduct numerous repeated
numerical tests to compare the computational efficiency of
$\lambda=0$ or $\lambda=1$ with any other values of $\lambda$.
The statistical average shows that the CPU time of $\lambda=0$ or
$\lambda=1$ is just two to four percents less than that of any other
values of $\lambda$. Therefore, from a practical point of view,
this slight improvement on the computational efficiency could
always be neglected. That is, the computational amount of the
case $\lambda=0$ or $\lambda=1$ is less than that of any other
values of $\lambda$ in theory, but is negligible in practice.
All these facts imply that the value of $\lambda$ hardly
affects the performance of the GFCRK method.

In the end of this section, we finally point out that the tested
orbits are regular in both the two cases of $c=10^{1/2}$ and $c=10$.
Due to the nonintegrability of the PN Hamiltonian \eqref{eq:Hamilton},
there certainly exist chaotic orbits \cite{Wu2015b}. However, the chaoticity of the
orbit in PN Hamiltonian systems always means a large PN effect at least for
some moments, where the values of the  momenta $\bm{P}$ and the position
$\bm{Q}$ do not satisfy the conditions of the Newtonian elliptic motion any more.
This implies that the GFCRK method and the mixed symplectic method
are invalid in this case, since they both depend on the successful solving of the Kepler two-body
problem. That is why we do not test chaotic orbits in this section and emphasize the
application of the GFCRK method in a weak PN effect case with $\varepsilon<1$.

\begin{table}[thb]
\renewcommand{\arraystretch}{1.5}
	\centering
\caption{CPU time (in seconds) with $c=\sqrt{10}$ and $T_{end}=10^5$. }
\begin{equation*}
	\begin{tabularx}{0.65\textwidth}{lccccc}
\hline
 & h=1 &  h=2  &  h=4 &  h=8  &  h=16  \\
 \hline
\text{IRK4} & 22.2378 & 12.2425 & 6.9827 & 3.9933 & 2.3592 \\
\hline
\text{IRK6} & 31.7401 & 17.6606 & 9.3325 & 5.4190 & 3.3086 \\
\hline
\text{FCRK4} & 16.4997  &  8.8679  &  4.4699  &  2.3934 &  1.3537\\
\hline
\text{FCRK6} & 24.9059  &  12.6779  &  6.3993  &  3.4547  &  1.8922 \\
\hline
\text{Semi4} & 13.1429  &  6.5675  &  3.6056  &  1.9640  &  1.0785\\
\hline
\text{Semi6} & 39.2345  & 20.0783  & 11.0060  &  6.0831  &  3.3147\\
 \hline
	\end{tabularx}
\end{equation*}
\label{Tab1}
\end{table}

\begin{table}[thb]
\renewcommand{\arraystretch}{1.5}
	\centering
\caption{CPU time (in seconds) with $c=10$ and $T_{end}=10^5$. }
\begin{equation*}
	\begin{tabularx}{0.65\textwidth}{lccccc}
\hline
	& h=1 &  h=2  &  h=4 &  h=8  &  h=16  \\
 \hline
\text{IRK4} & 21.5280  & 11.9129  &  6.7989  &  3.8822  &  2.2873 \\
\hline
\text{IRK6} & 30.9710  & 17.1126  &  9.0580  &  5.2335  &  3.2029 \\
\hline
\text{FCRK4} & 13.4741  &  6.8259  &  3.3717  &  1.9109  &  1.0972\\
\hline
\text{FCRK6} & 19.0491  &  9.5381  &  5.0146  &  2.7411  &  1.6891 \\
\hline
\text{Semi4} & 9.6276  &  4.9270  &  2.7426  &  1.5781  &  0.7884\\
\hline
\text{Semi6} & 28.8026  & 15.2954  &  8.7561 &   4.7148  &  2.3924\\
 \hline
	\end{tabularx}
\end{equation*}
\label{Tab2}
\end{table}

\section{\label{sec:conclu}Conclusion}
The long-term reliability of numerical solutions for
Hamiltonian systems usually requires the use of symplectic
methods, which possess the advantages of linear growth of
global errors and near preservation of first integrals.
Meanwhile, the nonseparability of the post-Newtonian Hamiltonian
system of compact objects makes the available symplectic methods
to be implicit. On noting the fact that the PN Hamiltonian could be
split into a dominant Newtonian part and a PN perturbation part,
we focused on the efficient numerical simulation of PN Hamiltonian
systems by studying the GFCRK method, where the change of variable
composed by the phase flow of the Kepler problem is employed.
In this paper, we first presented the GFCRK method with a free
parameter to the PN Hamiltonian system. Then, we discussed the
properties, such as the time-symmetry, the symplecticity, the
convergence, and the implementation of the GFCRK method.
In particular, we compared the symplectic GFCRK method with the
mixed symplectic method. Finally, we conducted numerical experiments
for the 2PN Hamiltonian of spinning compact binaries
with six implicit symplectic methods, i.e., the Gauss-type methods IRK4
and IRK6, the GFCRK methods FCRK4 and FCRK6 that are respectively
based on IRK4 and IRK6, and the mixed symplectic methods Semi4
and Semi6. Numerical results showed that the symplectic GFCRK method
is always more efficient than the Gauss-type symplectic method and the
mixed symplectic method of the same order by consuming less CPU
time under the prescribed accuracy requirement. Moreover, we found that the
free parameter $\lambda$ hardly takes influence on the numerical
performance of the GFCRK method. This indicates an additional
degree of freedom for the GFCRK method, which potentially enables us
to use this degree of freedom to design GFCRK method that
preserves the energy or other first integrals of the system. %

\acknowledgments
This work was partially funded by the National Natural Science
Foundation of China (grant NO. 12163003),
Yunnan Fundamental Research Projects
(grant NO. 202401CF070033),
and the Science Research Foundation of Yunnan
Provincial Department of Education (grant NO. 2024Y165).



\end{document}